\documentclass{article}

\pdfoutput=1

\usepackage{amsfonts}
\usepackage{array}
\usepackage[table,xcdraw]{xcolor}
\usepackage{booktabs}
\usepackage{fancyhdr}
\usepackage{orcidlink}
\usepackage{subcaption}
\usepackage{url}
\usepackage{graphicx}
\usepackage{hyperref}
\hypersetup{hidelinks}

\usepackage{PRIMEarxiv}

\title{Measuring software innovation with open source software development data}

\author{
  Eva Maxfield Brown \orcidlink{0000-0003-2564-0373} \thanks{Corresponding author: Eva Maxfield Brown, evamxb@uw.edu} \\
  University of Washington \\
  Seattle, WA, USA\\
   \And
  Cailean Osborne  \orcidlink{0000-0002-4018-8488} \\
  University of Oxford\\
  Oxford, UK\\
 \AND
  Peter Cihon \orcidlink{0009-0002-1540-0878} \\
  GitHub \\
  San Francisco, CA, USA\\
   \And
  Moritz Böhmecke-Schwafert \orcidlink{0000-0003-0543-2622} \\
  Technical University of Berlin \\
  Berlin, Germany\\
 \AND
  Kevin Xu \orcidlink{0009-0001-3987-1239} \\
  GitHub \\
  San Francisco, CA, USA\\
   \And
  Mirko Boehm \orcidlink{0000-0001-7658-3896} \\
  The Linux Foundation \\
  Berlin, Germany\\
   \AND
   Knut Blind \orcidlink{0000-0002-6510-122X} \\
  Technical University of Berlin, Fraunhofer ISI \\
  Berlin, Germany\\
}

\begin{document}
\maketitle
\vspace{-1em}

\begin{abstract}
Existing innovation metrics inadequately capture software innovation, creating blind spots for researchers and policymakers seeking to understand and foster technological innovation in an increasingly software-defined economy. This paper introduces a novel measure of software innovation based on open source software (OSS) development activity on GitHub. We examine the dependency growth and release complexity among $\sim$350,000 unique releases from 33,000 unique packages across the JavaScript, Python, and Ruby ecosystems over two years post-release. We find that the semantic versioning types of OSS releases exhibit ecosystem-specific and maturity-dependent patterns in predicting one-year dependency growth, with minor releases showing relatively consistent adoption across contexts while major and patch releases vary significantly by ecosystem and package size. In addition, while semantic versioning correlates with the technical complexity of the change-set, complexity itself shows minimal correlation with downstream adoption, suggesting that versioning signals rather than technical change drive dependency growth. Overall, while semantic versioning release information can be used as a unit of innovation in OSS development complementary to common sources for innovation metrics (e.g. scientific publications, patents, and standards), this measure should be weighted by ecosystem culture, package maturity, and release type to accurately capture innovation dynamics. We conclude with a discussion of the theoretical and practical implications of this novel measure of software innovation as well as future research directions.
\end{abstract}

\keywords{Open source software  \and innovation indicators \and GitHub \and dependency analysis \and semantic versioning \and release complexity \and software ecosystems}

\section{Introduction}
There is interdisciplinary research and policy interest in measuring innovation, with attention traditionally focusing on academic outputs \cite{dziallas2019innovation}, business outcomes \cite{oecd_oslo_2018}, and their interactions \cite{blind_interplay_2022}. Yet, software innovation, particularly open source software (OSS) innovation, often eludes conventional metrics such as scientific publications, patents, and standards. Although OSS are digital public goods that firms leverage as an ``inward knowledge flow'' to produce innovations \cite{oecd_oslo_2018}, it also constitutes innovation in its own right. The OECD Oslo Manual \cite{oecd_oslo_2018} frames innovation through product and process advances, evaluating novelty either by comparing to the current state of the art or by assessing ex-post impact. This framework poses unique challenges for OSS, where individual contributions may appear incremental---often just a few lines of code---but collectively drive radical innovation and creative destruction \cite{boehm_standard_2021}. This tension between granular contributions and systemic impact highlights the need for new approaches to measuring OSS innovation.

Previously, researchers have sought to measure OSS innovation through collaboration activity \cite{dabbish2012social,osborne_ai_2024}, its economic impact \cite{blind_impact_2021,wright-github-entrepreneurship}, and economic value \cite{hoffmann2024value,korkmaz2024github}. In this paper, we define and operationalize a novel method for measuring OSS innovation in discrete units, offering researchers and policymakers a software innovation metric that is complementary to traditional sources for innovation metrics. Our method uses publicly available data from GitHub, leveraging software developers’ publication of packages and other developers’ decisions to adopt the software. Developers often use semantic versioning when publishing a change to a software package that denotes its significance as major, minor, or  patch \cite{semverSemanticVersioning}. In practice, then, versioning increments may represent inventive steps in software. By reviewing other developers' activity in response to new versions, the open source community can offer external validation of whether an increment is, in fact, innovative. Specifically, we assess package dependencies by examining how frequently new versions are integrated into other projects over time, providing a metric to evaluate the impact of these inventive steps.

\textit{\textbf{Problem statement:}} Conventional innovation metrics inadequately capture the granular, iterative, yet collectively transformative nature of software innovation, creating blind spots for researchers, software managers, and policymakers seeking to understand and support technological innovation in an increasingly software-driven economy.

\textit{\textbf{Data \& methods:}  We analyze a dataset of approximately 350,000 unique releases from 33,000 unique packages across the JavaScript, Python, and Ruby ecosystems on GitHub, examining package dependencies and release complexity over time frames up to two years post-release. Our methodology combines quantitative analysis of dependency growth patterns with a machine learning approach to assess release complexity using GPT-4.}

\textit{\textbf{Key finding 1:} Semantic versioning types exhibit ecosystem-specific and pre-release utilization-dependent patterns in predicting dependency growth at both one-year and two-year post-release time frames. Minor releases demonstrate relatively consistent adoption benefits across JavaScript, Python, and Ruby ecosystems and initial dependency levels. In contrast, major and patch releases vary significantly based on ecosystem culture and baseline package popularity.}

\textit{\textbf{Key finding 2:} Ecosystem-specific adoption patterns remain consistent throughout all stages of a package’s development, from early (pre-1.0) to mature (2.0 and beyond). Across these stages, minor releases reliably achieve steady adoption, highlighting their role as innovation signals for OSS.}

\textit{\textbf{Key finding 3:} A release's technical complexity correlates with semantic versioning type but shows minimal direct correlation with dependency growth, indicating that versioning serves as a signaling mechanism that drives adoption independently of actual technical complexity.}

\textit{\textbf{Methodological implications:} These ecosystem-specific and context-dependent patterns suggest that future innovation measurement should normalize adoption metrics by ecosystem culture, baseline package popularity, and development lifecycle stage, rather than treating semantic versioning uniformly across contexts. Furthermore, by operationalizing dependency growth following semantic versioned releases as a scalable, real-time proxy for software innovation, our approach provides researchers, practitioners, and policymakers with a transparent and reproducible metric that complements traditional innovation indicators and enables more granular analysis of digital innovation dynamics }.

\textit{\textbf{Limitations:} We do not account for software licensing effects on dependency relationships (e.g. GPL vs. Apache v2), or  geographic patterns of adoption. These aspects were excluded due to limitations in the available data and the complexity of reliably linking licenses and geographic meta data to dependency events at scale. Additionally, we did not address the the inherent variability in our LLM-based semantic analysis methodology, as we prioritized scalability and automation over manual validation, considering the large data set. Future research could address these dimensions to further refine and contextualize software innovation metrics } 

\textit{\textbf{Impact statement:} Our findings provide new insights into OSS innovation dynamics, by revealing  ecosystem-sensitive metrics that capture the adoption and influence of OSS releases. This enables researchers to more accurately measure digital innovation, equips practitioners with data-driven tools for strategic decision-making, and empowers policymakers to design more effective interventions for fostering OSS-driven technological progress. }

The paper is structured as follows. In \S~\ref{sec:relatedwork}, we discuss prior work on innovation indicators and the gap for OSS innovation. In \S~\ref{sec:method}, we discuss our approach to measuring units of software innovation using GitHub data. In \S~\ref{sec:results}, we report our key findings. In \S~\ref{sec:discussion}, we discuss the implications and limitations of our method, as well as future research directions.

\section{Related Work} \label{sec:relatedwork}
This section reviews prior work on innovation indicators and discusses the absence of indicators for OSS innovation. We then establish a conceptual foundation for developing OSS innovation indicators by reviewing prior OSS measurement approaches and available data on OSS development and innovation. 

\subsection{Innovation indicators and the open source gap}
Innovation plays a fundamental role in driving long-term economic growth by fostering productivity improvements, technological advancements, and competitive advantages \cite{agh2014innovation}. According to the 2018 OECD Oslo Manual \cite{oecd_oslo_2018}, innovation refers to the  ``introduction of a new or significantly improved product or process that differs substantially from previous offerings''. For products, this involves being made available to potential users, while for processes, it means being brought into use by the organization. Consequently, the accurate and efficient measurement of innovation is essential for both policymakers and organizations \cite{oecd_oslo_2018}.

Innovation indicators are measurable values that provide insights into specific aspects of innovation or its overall status. As defined by the OECD \cite{oecd_oslo_2018}, these indicators summarize observed innovation phenomena within a population over time and enable the analysis of behaviors and activities, including developmental, financial, and commercial measures taken by organizations. The evaluation of innovation is complex, with various methodologies discussed in the literature. Dziallas and Blind (2019) \cite{dziallas2019innovation} identified 82 distinct indicators in their systematic review of innovation indicators. Based on this, we provide a summary of common innovation indicators in \textbf{Table~\ref{tab:innovation-indicators}}.

Innovation indicators constitute a fundamental metric for the evaluation of innovation performance. Companies utilize these metrics to allocate resources and evaluate the efficacy of innovation processes \cite{evanschitzky2012success,dewangan2014towards}. For policymakers, innovation indicators are indispensable in shaping and implementing mission-oriented innovation policy \cite{CanterVannuccini2018innopol}. They provide a means to rigorously evaluate regulatory impacts and the efficacy of funding initiatives \cite{bloom2019toolkit}, and facilitate a data-driven approach to policy development and resource distribution \cite{markianidouetal2022operational}.
Moreover, investors rely on innovation indicators to inform decisions on funding new ventures, underlining their significance for entrepreneurship \cite{dziallas2019innovation}. 

\begin{table}[h]
\centering
\caption{Characteristics of innovation output indicators adapted from Blind et al (2023) \cite{blind_et_al_2023_oss_inno}}
\label{tab:innovation-indicators}
\begin{tabular}{>{\raggedright\arraybackslash}p{2.5cm}>{\raggedright\arraybackslash}p{2.9cm}>{\raggedright\arraybackslash}p{2.9cm}>{\raggedright\arraybackslash}p{2.9cm}>{\raggedright\arraybackslash}p{2.9cm}}
\toprule
 & \textbf{Publications} & \textbf{Patents} & \textbf{Standards} & \textbf{Start-ups} \\
\midrule
\textbf{Input} & Scientific references (backward citations) & Patents, NPL references, standards & Standards, Scientific references & Patents, Standards, Open source \\
\addlinespace
\textbf{Process} & Co-authors plus affiliations & Co-inventor & Standardization working group (limited visibility) & Start-up teams \\
\addlinespace
\textbf{Assessment} & Peer review and acceptance by editors based on novelty and contribution & Granting by patent office based on novelty and inventive step & Publication based on stakeholder consensus and state of practice & Decision by founder/investors based on expected commercial success \\
\addlinespace
\textbf{Impact of output} & Forward citations in publications, patents, standards & Forward citations in patents, publications, standards & Citations in standards, patents, publications & Investments, employment \\
\addlinespace
\textbf{Measurement of implementation} &Indirectly via references in patents and standards & Patent-product database & Specific cases (e.g. laptops) or web mining & Revenue with product portfolio \\
\bottomrule
\end{tabular}
\end{table}

Kleinknecht et al. \cite{kleinknechtetal2022} argue that widely used innovation indicators, such as R\&D spending and patent applications, have significant limitations, including varying patent valuations and time lags. Consequently, using a broader range of data sources is essential for constructing more comprehensive indicators. The OECD criteria further emphasize this, focusing on four dimensions: Relevance, which ensures indicators meet user needs; Accuracy/Validity, requiring unbiased representations of innovation; Reliability/Precision, ensuring consistent results when repeated; and Timeliness, which stresses the availability of data for decision-making \cite{oecd_oslo_2018}.

Additionally, underpinning indicators include methodologies for measuring innovation, such as the selection of units of innovation, with novelty being a key consideration. The OECD defines novelty as a criterion for determining whether a product or process is ``significantly different'' \cite{oecd_oslo_2018} from previous iterations, thus qualifying as an innovation. This evaluation can be approached in two ways: first, by comparing a firm's innovations with the state-of-the-art within the firm and its market or industry; and second, by assessing the impact of an innovation in transforming or creating a new market, which may indicate radical or disruptive innovation. Unlike the first approach, assessing the impact of an innovation can only be done ex-post, often requiring several years to evaluate its effects.

Hence, the 2018 version of the OECD Oslo Manual advocates for the inclusion of novel digital data sources, emphasizing how automated data collection can expand the range of available data, thereby enhancing the development of more diverse and dynamic innovation indicators \cite{oecd_oslo_2018} \cite{böhmeckedörries2023}. Traditional indicators, in particular, often lack timeliness and accuracy and tend to underestimate innovation activities in some contexts \cite{böhmeckedörries2023}. In response, recent studies have begun to explore the potential of novel indicators, for example, based on web-based data sources such as startup databases and online repositories (see e.g. \cite{Kinne_Axenbeck_2020} \cite {böhmecke2023} \cite {böhmeckedörries2023}).

To date, OSS indicators are rarely discussed in prior innovation scholarship. For example, a systematic literature review of innovation indicator studies did not even mention ``open source'' \cite{dziallas2019innovation}. In 2018, the OECD considered OSS only under ``inward knowledge flows'', but did not consider it as innovation per se \cite{oecd_oslo_2018}. However, (open source) software development is an act of innovation in its own right, which contributes to product and process innovations alike that generate significant economic value. This is particularly true in the context of emerging technologies, where traditional metrics such as patents are increasingly inadequate for capturing the full scope of innovative activity \cite{böhmecke2023}. Hence, the rapid pace of development and the collaborative ethos inherent in OSS projects frequently result in innovations that elude conventional patenting processes, underscoring the necessity for alternative indicators to evaluate their impact.

Recently, research interest in the economics of OSS has begun to increase, albeit not for OSS innovation indicators directly. 
Hoffmann et al. (2024) \cite{hoffmann2024value} estimate the supply-side value of OSS at \$4.15 billion, while the demand-side value is estimated to reach a significantly larger \$8.8 trillion. The authors conclude that firms would need to spend 3.5 times more on software if OSS did not exist, highlighting not only its economic impact of OSS but also its innovative potential in software development. Meanwhile, Korkmaz et al. (2024) \cite{korkmaz2024github} develop a framework for measuring OSS value using GitHub data, estimating US investment in OSS at \$37.8 billion in 2019. Furthermore, Wright et al. (2023) \cite{wright-github-entrepreneurship} demonstrate a correlation between country-level GitHub participation and increased founding of innovative ventures, suggesting that OSS activity serves as a catalyst for innovation-driven entrepreneurship. Lin and Maruping (2021) \cite{lin2022open} add nuance to this relationship, showing that the benefits of OSS engagement for startups vary across their lifecycle stages, with different advantages accruing from inbound and outbound OSS activities.

The way in which OSS is developed presents challenges for the measurement of innovation, with tensions between micro- and macro-level effects of OSS development activity. At the micro-level, individual contributions to an OSS repository, such as a few lines of code in a commit, may be too marginal to qualify as an inventive step. By comparison, patents require a significant inventive step and a certain level of novelty to be granted. Yet, when considered at scale, the OSS community contributes to radical innovation and creative destruction in the ICT sector \cite{boehm_standard_2021} and OSS is used widely, making up 96\% of software stacks according to recent surveys \cite{synopsys_open_2023}. This underscores the discrepancy between the incremental contributions made by individuals at the micro-level and the substantial impact of OSS on innovation and its widespread incorporation into modern software stacks, an area that has yet to be thoroughly examined.

In addition, one must take into consideration that OSS represents a unique innovation model that blends elements of user innovation \cite{Hippel2001InnovationBU}, co-creation \cite{zwass_co-creation_2010}, and open science \cite{vonkrogh2007open}. This model is characterized by a complex ecosystem of contributors with diverse economic, social, and technological incentives, from individuals \cite{von2012carrots, gerosa2021shifting} to research institutes and universities \cite{cheng_open_2020} to multinational corporations \cite{li2024systematic,osborne2024-os-coopetition}. Companies---both large and small---use, contribute to, and fund OSS development \cite{blind_impact_2021,blind_estimating_2024,osborne2024public}. Thus, innovations in OSS may be the outcome of contributions across diverse organizations and developers. Due to this open collaboration model, value creation and value capture operate differently than in traditional R\&D processes or team structures in university or industry labs \cite{Bohm2022}.

Given the gap in OSS innovation indicators and the challenges presented by the nature of OSS development, there is a pressing need to develop new approaches for measuring units of innovation in this domain. The following sections explore approaches that have been proposed or may be adapted to address this challenge, drawing insights from (1) established innovation indicators not specifically related to OSS and (2) prior approaches to measuring OSS innovation using data from popular OSS hosting and development platforms such as GitHub. By examining these approaches, we aim to lay the groundwork for developing more comprehensive and precise indicators of OSS innovation that can capture both the incremental nature of individual contributions as well as the transformative potential of OSS at scale.

\subsection{Learning from existing innovation indicators}
In this section, we establish connections between established innovation indicators and potential OSS innovation indicators. 
Scientific publications offer a well-established benchmark for innovation measurement. Key performance indicators include the volume of output and forward citations received \cite{moed2009new}. Brown et al. (2023) demonstrated a similar approach for measuring the impact of scientific OSS repositories and software projects by creating large networks of scientific publications with links to software used and mentioned in each article \cite{Brown_Exploring_the_dependencies_2023}. This approach aligns with traditional bibliometric analyses but applies them to software artifacts. 

Patents provide another crucial external indicator. They can reference OSS projects, particularly in non-patent literature (NPL) citations, indicating the industrial relevance and potential commercial value of a repository. As Harhoff et al. (2003) \cite{harhoff2003citations} note, NPL citations are particularly important in fields such as pharmacy and chemistry, while in the ICT sector standards are also frequently cited \cite{bekkers2020impact}. For OSS repositories, one can track the number of patents citing them and analyze the characteristics of the citing patents and their applicants.

Standards are another established indicator of innovation \cite{blind_standardization_2019}. Standards play a crucial role in innovation diffusion, which is a measure of the success of innovative products and processes. Empirical evidence shows that involvement in standardization \cite{blind_standard-relevant_2022} and the implementation of standards \cite{mirtsch2020standards} significantly correlate with companies' innovation performance. For OSS repositories, we can examine the number of links to repositories in standards documents and analyze how these standards are subsequently used or cited. The European Commission's recent Code of Practice on Standardization (2023) \cite{eu2023codeofpractice} recognizes standards as outputs in the context of research projects, further validating their importance as innovation indicators. However, it is worth noting that contributor information for standards is often limited, with only a few standard-setting organizations, such as AFNOR in France, releasing this data \cite{hess2022coordination}.

Start-up ecosystems provide another indicator of OSS innovation. As Autio et al. \cite{autio2014entrepreneurial} argue, start-ups can be effective drivers for the implementation of innovation, playing a crucial role in economic development \cite{baumol2007entrepreneurship}, technological advancement, and societal impact \cite{wright-github-entrepreneurship}. For OSS repositories, we can analyze the use of repositories by start-ups, examining factors such as VC funding and growth rates of start-ups referencing specific repositories.

The implementation of innovations in practice is a critical aspect of their impact and success. While scientific publications may not be immediately implemented, they can indirectly influence practice through citations in patents \cite{sorenson2001science} or standards \cite{blind_standardization_2019}. For patents, recent research has begun to establish explicit links to specific products \cite{uhlbach_revealing_2023}. Standards implementation has been analyzed through case studies, such as Biddle et al.'s \cite{biddle2010many} examination of standards in laptops, and more recently through large-scale web mining approaches \cite{mirtsch2020standards}. 

Given the wealth of data available on developer platforms such as GitHub, which includes real-time metrics on repository usage and collaboration patterns, integrating indicators derived from these platforms with established innovation metrics can establish a robust framework for measuring OSS innovation.



\subsection{Prior approaches to measuring OSS innovation}

\begin{table}[t]
    \centering
    \small
    \caption{Characteristics of OSS repositories and comparison with innovation indicators}
    \label{tab:oss-indicators}
    
    \subfloat[Characteristics of OSS repositories\label{tab:oss-indicator-characteristics}]{
        \begin{tabular}{>{\raggedright\arraybackslash}p{4cm}>{\raggedright\arraybackslash}p{9cm}}
        \toprule
        \textbf{Indicators} & \textbf{Implications for Innovation} \\
        \midrule
        Commits & Input indicator; increasing commits suggest relevance and impact \\
        \addlinespace
        Stars & Weak indication of relevance, not necessarily innovativeness \\
        \addlinespace
        Watching & Indicator of relevance and impact, not direct innovation \\
        \addlinespace
        Forks & Potential for innovation through combination and modification \\
        \addlinespace
        Pull requests & Input indicator, not direct measure of innovation \\
        \addlinespace
        Dependencies & Revealed-preference view of project's influence and use \\
        \bottomrule
        \end{tabular}
    }
    
    \vspace{1em}
    
    \subfloat[Comparison between patents, standards, and OSS repositories adapted from  Blind et al (2023) \cite{blind_et_al_2023_oss_inno}\label{tab:comparison-patents-standards-oss}]{
        \begin{tabular}{>{\raggedright\arraybackslash}p{2.2cm}>{\raggedright\arraybackslash}p{2.8cm}>{\raggedright\arraybackslash}p{2.8cm}>{\raggedright\arraybackslash}p{2.8cm}>{\raggedright\arraybackslash}p{2.8cm}}
        \toprule
         & \textbf{Publications} & \textbf{Patents} & \textbf{Standards} & \textbf{Open Source Repositories} \\
        \midrule
        \textbf{Authors} & Co-authors & Inventors and applicants & Not available (exceptions) & Contributors \\
        \addlinespace
        \textbf{Contributions} & Not available (exceptions)& Not available (exceptions) & Not available (exceptions) & Commits or dependencies \\
        \addlinespace
        \textbf{Inputs to determine innovation} & Scientific papers, data & Patents, non-patent literature, scientific papers, standards & Standards, scientific papers, patents (incl. SEP) & Existing code in repositories \\
        \addlinespace
        \textbf{References to outputs} & Publications, patents, standards, OSS, products, start-ups & Patents, publications, standards, OSS, products, start-ups & Standards, publications, patents, products, start-ups & Other repositories, publications, patents, standards, products, start-ups \\
        \bottomrule
        \end{tabular}
    }
\end{table}

The measurement of OSS innovation builds upon diverse methodological traditions and data sources. Early work focused on adapting traditional innovation metrics to software contexts. Edison et al. (2013) \cite{edison2013towards} developed a conceptual framework incorporating patents \cite{acs2002patents} and other indicators to evaluate innovation determinants, inputs, outputs, and performance in the software industry. Similarly, Nirjar (2008) \cite{nirjar_innovations_2008} proposed an ``index of innovativeness'' for software SMEs, linking innovation capability to growth stages.

Recent studies have leveraged repository-level metrics to measure OSS innovation at the project or repository level \cite{blind_impact_2021,bao2019large,cosentino2017systematic,osborne_ai_2024}. Fang et al. (2024) \cite{fang2024novelty} examine atypical recombinations of software libraries, finding that higher levels of innovativeness correlate with higher GitHub star counts, suggesting a link between innovation and popularity. However, they also note that more innovative projects often involve smaller teams and face higher abandonment risks, highlighting the complex relationship between innovation and sustainability in OSS. 

Network analysis and user adoption patterns offer complementary approaches to measuring OSS innovation. Banks et al. \cite{banks2022measuring} demonstrate  a positive correlation between the centrality of Python packages in dependency networks and download statistics. In addition, download data provide signals of user adoption and growth patterns. For instance, Wiggins et al. \cite{wiggins_heartbeat_2009} propose a method to measure the size of an OSS project’s user base and the level of potential user interest that it generates based on daily download patterns, finding that active users rapidly download software after new releases, while potential new users download software irrespective of releases. However, adoption patterns vary across ecosystems and often exhibit significant lags, typically ranging from 3 months to a year \cite{decan2018evolution,scarsbrook2023typescriptsevolutionanalysisfeature,Zerouali2018technicallag,stringer2020technicallag}.

GitHub offers rich data for OSS innovation measurement. The platform provides various indicators for the performance and popularity of OSS repositories, including commits, stars \cite{borges_whats_2018}, watching \cite{sheoran2014github}, forks \cite{jiang_why_2017}, and pull requests \cite{batoun2023githubprs}. Project metrics, in particular repository forks, have been shown to correlate with usage \cite{vargas2024estimating}, offering a proxy variable for usage and adoption data. The GitHub dependency graphs illustrates the projects a repository depends on and those that depend on it, offering a network view of a project's influence and reliance in the broader software ecosystem \cite{banks2022measuring}. \textbf{Table~\ref{tab:oss-indicator-characteristics}} discusses the merits and limitations of these internal indicators, and \textbf{Table~\ref{tab:comparison-patents-standards-oss}} shows key variables for indicators from OSS repositories, patents, and standards.

\section{Methods \& Data} \label{sec:method}
This section outlines the methodology and data sources utilized in our study. \S~\ref{glossary} provides a glossary of key terms essential to understanding the technical aspects of software dependencies, releases, and versioning. \S~\ref{gh-innovation-graph} details the construction of our GitHub repository dataset and dependency analysis. \S~\ref{gh-complexity} describes our approach to analyzing the complexity of releases using LLM annotation.

\subsection{Glossary of Key Terms} \label{glossary}
This study utilizes core concepts from OSS development.

\paragraph{Dependencies and Dependents}
A \textit{dependent} is a software project that relies on other software modules (known as \textit{dependencies}). While software projects often form complex networks of dependencies, our analysis focuses solely on immediate, first-order relationships between dependents and dependencies.

\paragraph{Software Releases}
A \textit{release} represents the official distribution of a software version, typically comprising:
\begin{itemize}
    \item A packaged version of the project's code
    \item Release notes documenting changes and new features
    \item Installation and integration instructions
\end{itemize}

\paragraph{Semantic Versioning}
Modern software releases follow a three-part numerical scheme \cite{semverSemanticVersioning}:
\begin{itemize}
    \item \textit{Major version} (first number): Significant changes that may break backward compatibility
    \item \textit{Minor version} (second number): New features that maintain compatibility
    \item \textit{Patch version} (third number): Bug fixes and incremental improvements
\end{itemize}

\paragraph{Early-Stage Projects}
Projects in early development phases often use a special versioning approach:
\begin{itemize}
    \item \textit{Zero-major} releases: When the major version remains at 0 during prototyping
    \item \textit{Zero-minor} releases: Minor and patch changes during early development
    \item These early-stage releases may see different adoption patterns compared to mature projects
\end{itemize}

\subsection{Analysis of GitHub Repository Data and Dependencies} \label{gh-innovation-graph}
\subsubsection{Data Generation} 
We constructed our dataset from internal GitHub databases. Most of the data, such as repository stars, forks, and programming languages, can be accessed through the GitHub API. Additionally, dependent data can be sourced from third-party packages, such as \url{https://github.com/nvuillam/github-dependents-info} or from services such as \url{https://ecosyste.ms/}. However, results from these third-party options may vary in dependent counts due to differing counting criteria and the absence of look-ahead or look-behind metrics, which are essential for comprehensive analysis reproduction. While public sources support future use of our method, to our knowledge, no third party has attempted a comprehensive collection of daily snapshots of repository metadata and dependency graph information for the time period considered in our study. Our dataset, available at \textbf{\url{https://github.com/evamaxfield/dg-uoi-analysis}}, is the combination of three datasets of innovation-related activities on the GitHub platform.

The first dataset includes \textit{repository information}, which contains details such as the repository's description, topics, number of stars, and number of forks. This data provides insight into the popularity and engagement levels of various repositories (see \textbf{Table~\ref{tab:oss-indicator-characteristics}}).

The second dataset focuses on \textit{package releases}. This data provides key information about the updated versions of packages, with each entry including a repository identifier alongside release notes describing changes for each version. For instance, an entry includes the account name, the repository name, the release version of a Python package with a specific name, and the corresponding date. For example, "my-account/my-repo" refers to the account and repository name, while "v1.2.3" indicates the package version, and the release date is September 25, 2024. This information is detected and logged daily through GitHub's automatic parsing of standard packaging files, such as pyproject.tom and setup.py for Python packages, and package.json for JavaScript packages.

The third dataset focuses on \textit{dependents of a package}, which identifies the downstream repositories that rely on specific packages. Each entry in this dataset includes a repository identifier for the dependent repository and the name of the package on which it depends. For instance, an entry indicates that "another-account/another-repo" depends on a Python package named "zyx" as of September 25, 2024. Similarly, this information is detected and logged daily via GitHub's automatic parsing of standard packaging files (see above).

We combine and filter these datasets to create a comprehensive view of package releases and their dependencies. First, we join each package release with the corresponding repository information for each day of newly detected package releases. Next, we implement a filtering mechanism to enhance data quality. Specifically, we remove package releases originating from repositories that do not have at least one star or are classified as ``forked repositories''. This step is crucial to eliminate packages that may be either extremely new, lacking community support, have not undergone sufficient development, or may be low-quality or automatically generated with little meaningful content (e.g. spam). This ensures that our dataset reflects more established and relevant packages, focusing on those with active development and community engagement. We apply another filter to exclude package releases that do not adhere to the \textit{semantic versioning specification} (see \S~\ref{glossary} above). During this process, we decompose the version string and categorize releases accordingly to major, minor, and patch. This classification allows us to systematically analyze the frequency and impact of updates based on the significance of the changes.
  
Next, we filter to ensure that only package releases from repositories whose names align with their respective package names are included in our dataset. Identifying the primary package for a repository is crucial because it simplifies understanding package dependencies and updates. For instance, the repository \url{https://github.com/pandas-dev/pandas} contains the source code for the widely used Python package ``pandas,'' thus, its package releases are included in our dataset. Conversely, if the repository were named ``pandas-dev/pandas-subpkgs,'' where the repository name (``pandas-subpkgs'') does not match the package name (``pandas''), those releases would be excluded. This filtering criterion helps avoid confusion when multiple packages are housed within a single repository, making it challenging to track dependencies accurately and understand which package is being referred to in analyses. By applying this criterion, we aim to focus on the packages directly associated with their respective repositories, thereby enhancing the quality and reliability of our dataset. 

Lastly, we enrich our dataset by calculating several measures. First, we integrate look-ahead metrics regarding repository stars and forks. This involves systematically retrieving and compiling repository data for each specified look-ahead day interval per repository. Specifically, for each repository, we collect data on the number of stars and forks over the defined intervals, which serves as an indicator of community engagement and interest. Second, we integrate the number of package dependents. This is accomplished by counting the instances of non-forked repositories that have received at least one star and include a dependency on a package sharing the same name, and within the same ecosystem (e.g. Python, JavaScript). 

 
\subsubsection{Data Processing}
Our dataset, compiled from dependency graph package release detection (see \S~\ref{gh-inno-graph-analysis}, in total includes 1,492,270 software releases from 167,375 unique packages from multiple different software ecosystems on GitHub. Among these, the most common ecosystems are JavaScript (NPM), Python (PyPI), and Ruby (RubyGems). 

Following dataset compilation, we implemented a series of necessary data processing \footnote{All of these pre-processing activities were executed using the Polars library in Python \cite{polars}, which facilitated efficient manipulation of large datasets and streamlined the data processing workflow} steps to prepare the data for analysis. First, we removed any packages with multiple releases on the same day. While it is common for packages to have multiple releases over longer time horizons, excluding same-day versions minimizes the risk of applying the same set of metrics to different releases.\footnote{This scenario typically arises in two contexts: (i) a new major or minor release of a package which may be followed by a patch release to rectify bugs inadvertently introduced with new features, and (ii) mature software projects may maintain multiple versions of the same software, such as a "Long-Term Support" (LTS) version alongside a leading-edge version.} Second, we excluded packages not belonging to the JavaScript, Python, and Ruby ecosystems, as these languages consistently accounted for at least 5,000 package releases across all datasets (spanning six-month, one-year, and two-year intervals). Third, we excluded any packages that did not have at least five dependents before the release. This criterion ensures that only actively utilized packages are included in our analysis, enhancing the focus on software that has demonstrated relevance within the ecosystem.

Following these steps, our filtered dataset consists of 356,220 unique releases from 33,236 unique packages from the JavaScript (NPM), Python (PyPI), and Ruby (RubyGems) ecosystems.\footnote{Release and package counts are based on the produced one-year look-ahead metric dataset. Different look-ahead time-span datasets resulted in different counts; however, we focused on the final counts for the one-year look-ahead metrics because our analysis also focuses on the same portion of the data.}

Subsequently, we stratified the dataset based on package popularity, measured by existing dependents. We categorized the packages into four log-scale bins: packages with less than 100 dependents (small), packages with between 100 and 1,000 dependents (medium), packages with between 1,000 and 10,000 dependents (large), and packages with greater than 10,000 dependents (huge). Additionally, we classified the packages by release series based on their versioning: zero-version (i.e., 0.*), one-version (i.e., 1.*), and two-plus-version (i.e., 2+.*). This dual stratification reflects both the popularity and maturity of the packages.

Finally, we calculated the log differences between each metric at designated look-ahead points and the corresponding release date. For instance, to compute the 90-day log difference in dependents following a package release, we subtracted the natural logarithm of the dependents on the release date from the natural logarithm of the dependents at 90 days post-release. This log difference calculation was performed for each metric (stars, forks, and dependents) across all specified look-ahead intervals: six months (with measurements every 45 days), one year (every 90 days), and two years (every 180 days).

\subsubsection{Data Analysis} \label{gh-inno-graph-analysis}
Our data analysis focused on examining the characteristics of the generated dataset and assessing the impact of software releases on subsequent software utilization. The following steps outline our approach.

\begin{figure}
    \centering
    \includegraphics[width=0.8\linewidth]{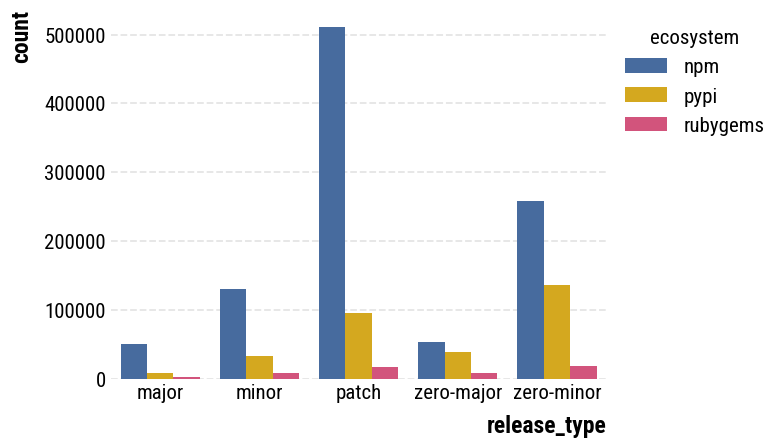}
    \caption{Release Counts by Release Type (Major, Minor, Patch, Zero-Major, and Zero-Minor and Ecosystem (NPM: JavaScript, PyPI: Python, RubyGems: Ruby)}
    \label{fig:ecosystem-release-type-release-counts}
\end{figure}

We began by analyzing the demographics of the dataset, aiming to understand its composition in terms of both programming ecosystems (i.e., JavaScript, Python, and Ruby) and the types of software releases. This demographic breakdown provided essential context for subsequent analyses by illuminating the distribution of releases across ecosystems and their respective release categories. When focusing on the NPM, PyPI, and RubyGems, we observe that a majority of our data comes from NPM (JavaScript) package releases. However, each ecosystem exhibits similar proportions for each release type which include major, minor, zero-major, and zero-minor releases (see \textbf{Figure \ref{fig:ecosystem-release-type-release-counts}}).

To explore the software releases' effect on future utilization, we calculated log-differences in dependents from the release date to one year after release. We select a one-year period because it is typically sufficient for users to integrate new features, adapt to changes, and evaluate implications of a release on their project \cite{decan2018evolution,scarsbrook2023typescriptsevolutionanalysisfeature,Zerouali2018technicallag,stringer2020technicallag}. These log-differences are a measure of how much the software's overall dependency network expanded during this period. We categorized these measurements by release type—major, minor, and patch—and conducted an Analysis of Variance (ANOVA) to determine whether there were statistically significant differences between groups. We also performed post-hoc pairwise Welch's independent t-tests to identify which specific release types exhibited significantly different distributions in the one-year change in dependents. The ANOVA and t-tests were executed using the SciPy library \cite{scipy}.

Next, we compared log-differences in dependents for prevalent ecosystems in our dataset: JavaScript, Python, and Ruby. This comparison allowed us to observe variations in software utilization across these dominant programming ecosystems for understanding ecosystem-specific behaviors. Further stratification of the dataset was conducted to investigate how dependent growth changes when controlling for version series (i.e., zero-version, one-version, and two-plus-version) and initial software utilization. For this, we divided the initial number of dependents into four groups: less than 100 dependents (small), at least 100 but less than 1,000 dependents (medium), at least 1,000 but less than 10,000 dependents (large), and at least 10,000 dependents (huge). We employed these subgroup analyses to reveal how software maturity and early-stage adoption influence long-term dependency growth. Data visualizations were generated using seaborn \cite{seaborn}, matplotlib \cite{matplotlib}, opinionated \cite{opinionated}, and colormaps \cite{colormaps}.

\subsection{Analysis of GitHub Releases and LLM Annotation of Complexity} \label{gh-complexity}
\subsubsection{Data Generation}
We closely follow the methodology established in \S~\ref{gh-innovation-graph} for the analysis of GitHub releases and complexity. We combine multiple private daily snapshot datasets encompassing repository information, package releases, and dependency details, ensuring consistency through uniform filtering and processing techniques. We perform two more steps to enrich our dataset. First, we focus on package releases that include an accompanying GitHub release, which allows us to capture the associated release notes. Second, we implement a stringent filtering criterion, removing releases lacking a minimum of 512 text characters in their release notes. This requirement ensures that the releases in our dataset contain sufficient textual information, enabling both our prompted LLM and domain experts to effectively assess the complexity of each release. These notes potentially provide critical context and insights into the changes made in each release.

Aside from these additional steps, the fundamental data generation process remains unchanged from that described in \S~\ref{gh-innovation-graph}, allowing for a seamless integration of our augmented data approach and maintaining methodological rigor.

\subsubsection{Data Processing}
Our data processing methodology involved a novel and structured approach to assessing the complexity of GitHub releases. After compiling a dataset of release notes and repository metadata, we rated the complexity of each release to better understand the effort required to implement the listed changes.

To achieve this, we employed a prompted GPT-4 model, which was iteratively refined to ensure accurate and consistent complexity ratings. The prompt, detailed in Appendix~\ref{sec:appendix-prompt}, adhered to best practices in prompt engineering, incorporating clear task definitions, example input-output pairs, and a defined structure for input and response \cite{lin2024write}. The model’s responses were formatted in XML, facilitating structured data extraction for subsequent analysis. This ensured consistency in how the model evaluated the complexity of each GitHub Release.

The complexity ratings were designed to reflect the difficulty a core developer of the project might face in implementing the changes described in the release notes. We defined complexity on a 7-point scale, ranging from 1 (minimally complex, requiring minimal changes and low expertise) to 7 (maximally complex, involving substantial changes and significant expertise). This scale allowed us to capture variations in both the number of changes and the depth of knowledge required to implement them.

Given that this complexity-scoring approach is novel, we thoroughly examined its robustness. To validate the model’s output, we compared its ratings with those provided by professional developers. A JavaScript developer evaluated the complexity of releases from JavaScript projects, while a Python developer did the same for Python projects. This human annotation process served as a benchmark to assess the accuracy of the GPT-4 model’s ratings. Our analysis revealed a positive correlation between the model and the developers’ assessments, with a correlation coefficient of $\alpha$ = 0.39 for JavaScript (p $<$0.00001) and $\alpha$ = 0.36 for Python (p $<$0.001). Notably, more than 60\% of the GPT-4 and developer ratings were within one complexity rank of each other (JavaScript: 68.5\%, Python: 63.3\%), indicating substantial alignment between the model and human experts.

However, a more detailed analysis of the complexity ratings distributions shows a divergence between the GPT-4 model and the professional developers, with the GPT-4 model consistently assigning higher complexity scores. GPT-4 consistently assigns a score of 4 or higher to releases that professionals rate as 3, especially at the 25th and 50th percentiles of complexity, overestimating the complexity of releases that professionals consider less challenging. This discrepancy may stem from the model’s sensitivity to detailed technical language or broader interpretation of the task’s complexity, leading to a higher baseline rating for changes it encounters.

Despite this upward bias, the model demonstrated substantial agreement with the professional developers' ratings within one complexity rank, particularly at the median and higher complexity levels. Given this high level of concordance, we used the finalized GPT-4 prompt to generate complexity ratings for the entire dataset of GitHub Releases and their corresponding release notes. This approach allowed us to efficiently scale the complexity assessment process while maintaining reasonable accuracy and alignment with expert human judgment.

\subsubsection{Data Analysis}\label{complexity-data-analysis}
Building upon the methodology described in \S~\ref{gh-inno-graph-analysis}, we employed a structured approach to investigate the characteristics and impact of complexity ratings across different programming ecosystems.

We began by examining the descriptive statistics of the dataset, specifically focusing on the distributions of complexity ratings across different release types (i.e., major, minor, and patch) and ecosystems (JavaScript, Python, and Ruby). This initial step helped establish a comparative understanding of how complexity ratings varied by release type and ecosystem. \textbf{Table \ref{tab:complexity-rating-descriptive-stats}} summarizes the descriptive statistics of the dataset.

\begin{table}[t]
\centering
\caption{Descriptive Statistics for Release Notes Complexity Ratings Across each Programming Language}
\label{tab:complexity-rating-descriptive-stats}
\begin{tabular}{lrrrrrr}
\toprule
\textbf{Prog. Lang.} & \textbf{\# Releases} & \textbf{Mean} & \textbf{StD} & \textbf{25\%} & \textbf{50\%} & \textbf{75\%} \\
\midrule
Python & 12186 & 4.52 & 1.16 & 4 & 5 & 5 \\
JavaScript & 11469 & 3.90 & 1.28 & 3 & 4 & 5 \\
Ruby & 2606 & 4.27 & 1.08 & 4 & 4 & 5 \\
\bottomrule
\end{tabular}
\end{table}

We conducted an ANOVA test to evaluate differences in the distributions of complexity ratings between ecosystems. We then performed post-hoc pairwise Welch's independent t-tests to identify where significant differences occurred between specific ecosystems. Finally, we sought to understand how complexity ratings were related to the subsequent software utilization by calculating the Spearman correlation ($\rho$) between the complexity rating of each release and the one-year log-difference in dependents. This analysis was performed separately for each major ecosystem in our dataset. The goal was to assess whether more complex releases tend to result in greater or lesser growth in the number of dependents over time, providing insight into how release complexity may influence the broader software ecosystem's uptake and reliance on the software in question.

\section{Results} \label{sec:results}
In this section, we present the results of our analysis. We begin by reporting the results about dependency measures in \S~\ref{gh-dep-results}. Subsequently, we report the results on release complexity as annotated by LLMs in \S~\ref{gh-complex-results}.

\subsection{Repository metrics and package dependencies} \label{gh-dep-results}

\subsubsection{Release adoption patterns are ecosystem and pre-existing utilization specific}

\textit{\textbf{Key finding:} Semantic versioning types exhibit ecosystem-specific and pre-release utilization-dependent patterns in predicting dependency growth at both one-year and two-year post-release timeframes. Minor releases demonstrate relatively consistent adoption benefits across JavaScript, Python, and Ruby ecosystems and initial dependency levels. In contrast, major and patch releases vary significantly based on ecosystem culture and baseline package popularity.}
 
Across the studied ecosystems and package sizes, the impact of release type on the log-difference in direct dependents one year after release reveals nuanced patterns (Table \ref{tab:one-year-log-diff-dependents-mean-std} and Figure \ref{fig:one-year-mean-log-diff-dependents}, see Appendix \ref{sec:appendix-one-year-timepoints-boxplots} and \ref{sec:appendix-two-year-timepoints-boxplots} for multiple-timepoint distributions over one and two years post-release). In the NPM ecosystem, we observe that major and minor releases tend to correlate with the largest average increase in log-difference for small, medium, and large-sized packages, respectively.

The PyPI ecosystem exhibits a different dynamic. Small PyPI packages show the largest average growth in direct dependents following patch releases, while medium and large-sized packages see comparable increases across all release types. Interestingly, major and minor releases are associated with the most substantial log-difference increase for huge PyPI packages.

In contrast, the RubyGems ecosystem shows a generally muted response to new releases regarding dependent growth. While small and medium-sized packages show slightly higher average log-differences after major and minor releases, respectively, the overall increases are not statistically significantly different. Notably, the largest RubyGems packages show virtually no positive change or slight negative trends for major and patch releases. This may reflect broader trends in the adoption and utilization of the Ruby ecosystem during the study period.

\begin{table}[]
\centering
\caption{Mean and Standard Deviation of Log-Difference in Direct Dependents One Year After Release, Stratified by Ecosystem, Package Size, and Release Type. Values in bold indicate the statistically significantly highest mean log-difference within each ecosystem and package size category when comparing major, minor, and patch releases (pairwise t-tests).}
\label{tab:one-year-log-diff-dependents-mean-std}
\resizebox{\textwidth}{!}{%
\rowcolors{2}{white}{gray!8} 
\renewcommand{\arraystretch}{1.3} 
\begin{tabular}{@{}llrrr@{}}
\toprule
Ecosystem & Package Size & \multicolumn{1}{l}{Log Diff. Major} & \multicolumn{1}{l}{Log Diff. Minor} & \multicolumn{1}{l}{Log Diff. Patch} \\ \midrule
npm & small & {0.104 $\pm$ 0.422} & {\textbf{0.12 $\pm$ 0.368}} & {0.107 $\pm$ 0.384} \\
& medium & {\textbf{0.117 $\pm$ 0.336}} & {0.103 $\pm$ 0.253} & {0.103 $\pm$ 0.246} \\
& large & {\textbf{0.104 $\pm$ 0.209}} & {0.081 $\pm$ 0.176} & {0.09 $\pm$ 0.18} \\
& huge & {0.042 $\pm$ 0.073} & {0.044 $\pm$ 0.098} & {0.044 $\pm$ 0.094} \\
\midrule 
pypi & small & {0.07 $\pm$ 0.226} & {0.088 $\pm$ 0.242} & {\textbf{0.114 $\pm$ 0.273}} \\
& medium & {0.088 $\pm$ 0.144} & {0.092 $\pm$ 0.135} & {0.088 $\pm$ 0.143} \\
& large & {0.063 $\pm$ 0.094} & {0.076 $\pm$ 0.107} & {0.069 $\pm$ 0.101} \\
& huge & {\textbf{0.043 $\pm$ 0.04}} & {\textbf{0.044 $\pm$ 0.038}} & {0.033 $\pm$ 0.017} \\
\midrule 
rubygems & small & {0.087 $\pm$ 0.399} & {0.081 $\pm$ 0.333} & {0.063 $\pm$ 0.315} \\
& medium & {0.05 $\pm$ 0.168} & {0.065 $\pm$ 0.186} & {0.059 $\pm$ 0.204} \\
& large & {0.012 $\pm$ 0.063} & {\textbf{0.023 $\pm$ 0.067}} & {0.015 $\pm$ 0.046} \\
& huge & {-0.002 $\pm$ 0.017} & {0.0 $\pm$ 0.017} & {-0.001 $\pm$ 0.019} \\ \bottomrule
\end{tabular}%
}
\end{table}

\begin{figure}[htbp]
    \centering
    \includegraphics[width=\textwidth]{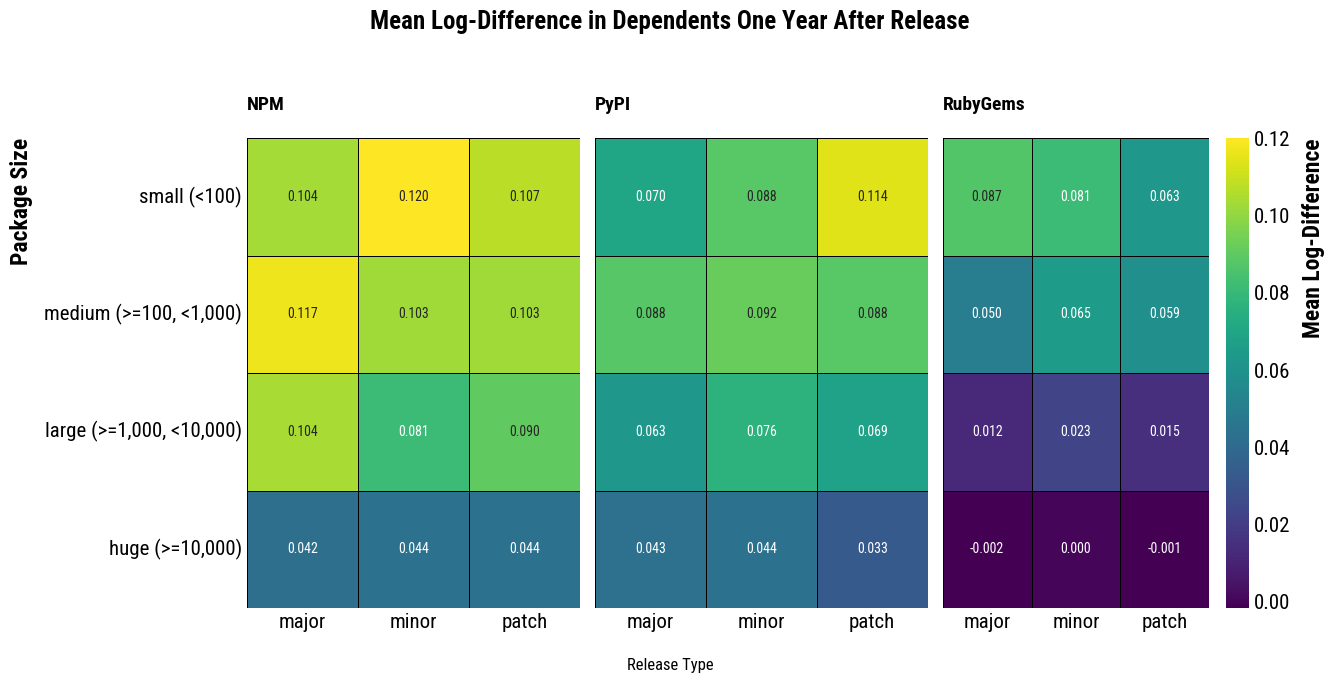}
    \caption{\textbf{Mean Log-Difference in Direct Dependents One Year After Release Across Ecosystems and Package Sizes.} This set of heatmaps illustrates the average log-difference in direct dependents for packages in the NPM, PyPI, and RubyGems ecosystems, categorized by package size (small, medium, large, huge) and release type (Major, Minor, Patch). The color intensity, consistent across all subplots, represents the magnitude of the mean log-difference, allowing for direct visual comparison of growth patterns across different ecosystems and package sizes following a new version release.}
    \label{fig:one-year-mean-log-diff-dependents}
\end{figure}

These findings suggest that the impact of a new package version on its adoption, as measured by the growth in direct dependents, is not uniform across ecosystems or package sizes. This variability highlights the importance of considering these factors when analyzing package impact, innovation, and ecosystem health.

\subsubsection{Release adoption patterns are ecosystem and version-series specific}
\textit{ \textbf{Key finding:} Analysis across versioning series maturity (zero-ver, one-ver, two-plus-ver packages) reveals that ecosystem-specific adoption patterns persist regardless of development lifecycle stage, with minor releases consistently performing as a stable middle ground across all versioning series, reinforcing their role as reliable innovation signals.}
 
Building on our understanding of how pre-existing package utilization influences adoption, we now examine the interplay between semantic versioning types and a package's version series maturity (zero-ver, one-ver, and two-plus-ver) in predicting dependent growth. This analysis, detailed in Table \ref{tab:one-year-log-diff-dependents-mean-std-ver-series} and Figure \ref{fig:one-year-mean-log-diff-dependents-ver-series} (See Appendix \ref{sec:appendix-one-year-timepoints-boxplots-ver-series} and \ref{sec:appendix-two-year-timepoints-boxplots-ver-series} for multiple-timepoint distributions over one and two years post-release, indicates that ecosystem-specific adoption patterns largely persist regardless of a package's development lifecycle stage. In the NPM ecosystem, major releases correlate with increased mean log-difference in direct dependents for both zero-ver (0.219 $\pm$ 0.525) and one-ver (0.281 $\pm$ 0.724) packages. For two-plus-ver NPM packages, mean log-difference values across major, minor, and patch releases are comparable.

The PyPI ecosystem presents a different dynamic. Major releases show a higher mean log-difference for zero-ver packages (0.101 $\pm$ 0.247). One-ver PyPI packages show comparable growth across release types. For two-plus-ver PyPI packages, patch releases show a higher mean log-difference (0.123 $\pm$ 0.265). The RubyGems ecosystem, consistent with previous findings, generally exhibits lower dependent growth across all version series. Minor releases show a higher mean log-difference for zero-ver RubyGems packages (0.15 $\pm$ 0.435), while major releases are higher for one-ver packages (0.136 $\pm$ 0.526). Differences in dependent growth across release types are minimal and low for two-plus-ver RubyGems packages.

\begin{table}[]
\centering
\caption{Mean and Standard Deviation of Log-Difference in Direct Dependents One Year After Release, Stratified by Ecosystem, Version Series, and Release Type. Values in bold indicate the statistically significantly highest mean log-difference within each ecosystem and package size category when comparing major, minor, and patch releases (pairwise t-tests).}
\label{tab:one-year-log-diff-dependents-mean-std-ver-series}
\resizebox{\textwidth}{!}{%
\rowcolors{2}{white}{gray!8} 
\renewcommand{\arraystretch}{1.3} 
\begin{tabular}{@{}llrrr@{}}
\toprule
Ecosystem & Version Series & Log Diff. Major & Log Diff. Minor & Log Diff. Patch \\
\midrule
npm & zero-ver & {\textbf{0.219 $\pm$ 0.525}} & {0.204 $\pm$ 0.549} & \\
 & one-ver & {\textbf{0.281 $\pm$ 0.724}} & {0.152 $\pm$ 0.396} & {0.136 $\pm$ 0.416} \\
 & two-plus-ver & {0.086 $\pm$ 0.32} & {0.092 $\pm$ 0.285} & {0.089 $\pm$ 0.304} \\
\midrule 
pypi & zero-ver & {\textbf{0.101 $\pm$ 0.247}} & {0.086 $\pm$ 0.259} & \\
 & one-ver & {0.099 $\pm$ 0.266} & {0.091 $\pm$ 0.237} & {0.08 $\pm$ 0.224} \\
 & two-plus-ver & {0.065 $\pm$ 0.191} & {0.086 $\pm$ 0.209} & {\textbf{0.123 $\pm$ 0.265}} \\
\midrule 
rubygems & zero-ver & {0.125 $\pm$ 0.399} & {\textbf{0.15 $\pm$ 0.435}} & \\
 & one-ver & {\textbf{0.136 $\pm$ 0.526}} & {0.07 $\pm$ 0.288} & {0.061 $\pm$ 0.317} \\
 & two-plus-ver & {0.05 $\pm$ 0.257} & {0.06 $\pm$ 0.254} & {0.047 $\pm$ 0.22} \\
\bottomrule
\end{tabular}%
}
\end{table}

\begin{figure}[htbp]
    \centering
    \includegraphics[width=\textwidth]{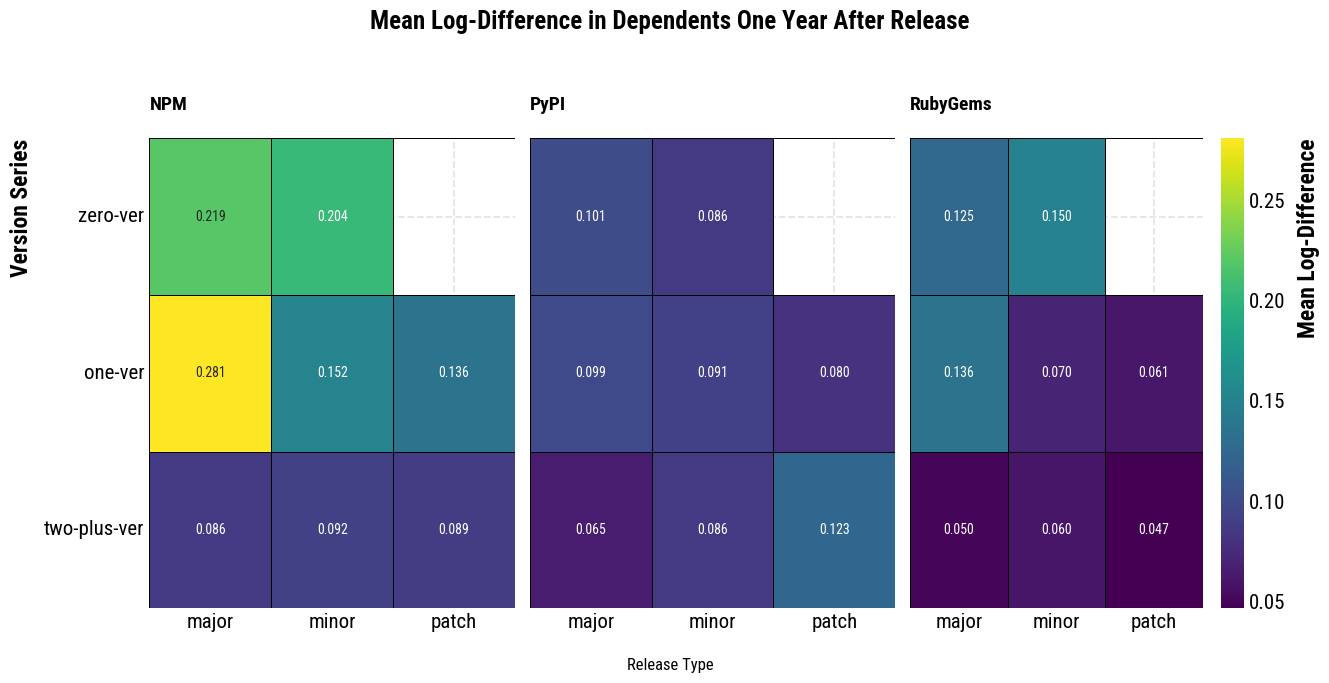}
    \caption{\textbf{Mean Log-Difference in Direct Dependents One Year After Release Across Ecosystems and Package Sizes.} This set of heatmaps illustrates the average log-difference in direct dependents for packages in the NPM, PyPI, and RubyGems ecosystems, categorized by version series (zero-version, one-version, two-plus-version) and release type (Major, Minor, Patch). The color intensity, consistent across all subplots, represents the magnitude of the mean log-difference, allowing for direct visual comparison of growth patterns across different ecosystems and version series following a new version release.}
    \label{fig:one-year-mean-log-diff-dependents-ver-series}
\end{figure}

These results indicate that semantic versioning's influence on package adoption remains connected to ecosystem culture, even as packages mature.

\subsection{LLM Annotations of Release Complexity} \label{gh-complex-results}

\subsubsection{Correlation between release complexity and dependency growth}

\textit{\textbf{Key finding:} A release’s technical complexity correlates with semantic versioning type but shows minimal direct correlation with dependency growth, indicating that versioning serves as a signaling mechanism that drives adoption independently of actual technical complexity}

\begin{figure}[htbp]
  \centering  
  
  \subfloat[Complexity Rating Distributions by Release Type (Major, Minor, or Patch)]
  {
    \includegraphics[width=0.8\linewidth, height=0.35\textheight]{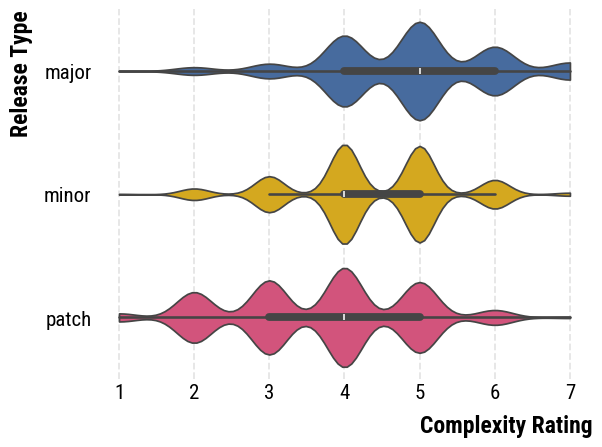}
    \label{fig:complexity-ratings-all-langs}
  }
  
  \vspace{5em}
 
  \subfloat[Complexity Ratings of Release Notes Correlate with Log-Difference of Dependents After One Year]
  {
    \includegraphics[width=0.9\linewidth, height=0.25\textheight]{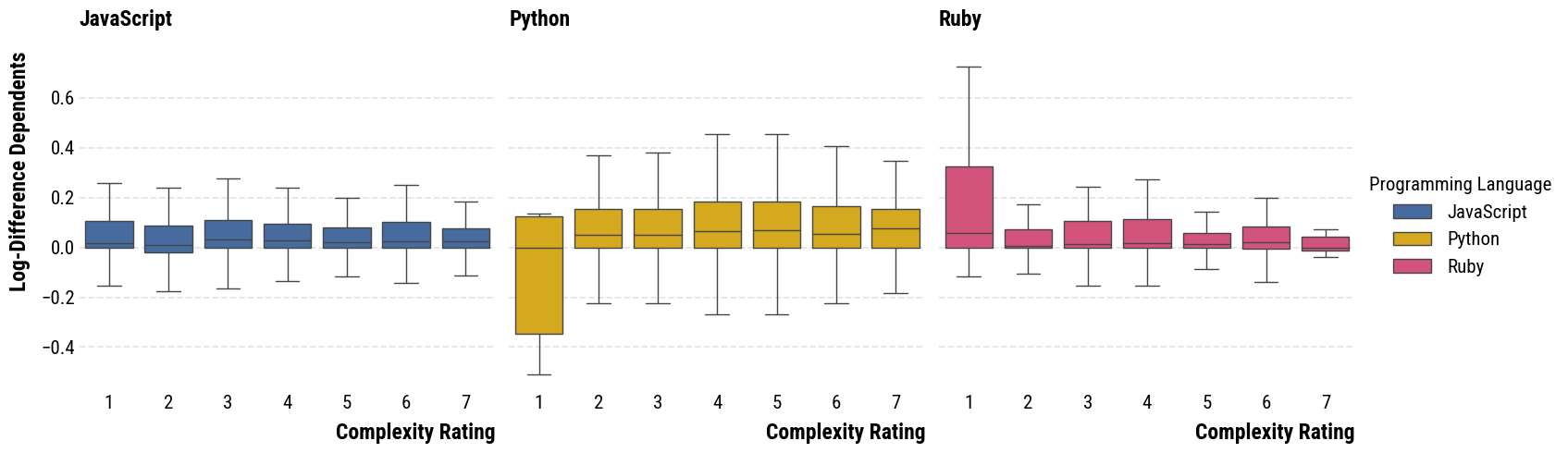}
    \label{fig:log-diff-dependents-release-complexities}
  }
  
  \caption{Complexity Ratings of Release Notes}
  \label{fig:complexity-ratings}
\end{figure}

Next, we investigated the "technical complexity to implement the changes" within a release, as rated by GPT-4, to understand how release complexity relates to semantic versioning types and subsequent package adoption. Our analysis, presented in Table \ref{tab:release-type-complexity-p-values} (and Figure \ref{fig:complexity-ratings-all-langs}), indicates a consistent pattern across JavaScript, Python, and Ruby: Major releases exhibit higher mean complexity ratings than Minor releases, which have higher mean ratings than Patch releases. For instance, in JavaScript, Major releases had a mean complexity of 4.71 ($\pm$1.10), Minor releases 4.14 ($\pm$1.00), and Patch releases 3.39 ($\pm$1.17). All pairwise comparisons between each language's release types show statistically significant differences (p $<$0.0001). This correlation between semantic versioning type and assessed release complexity suggests that maintainers generally adhere to semantic versioning conventions regarding the scope and effort involved in their releases.

\begin{table}[t]
\centering
\caption{Programming Languages, Release Types, and Complexity Rating Comparisons}
\label{tab:release-type-complexity-p-values}
\renewcommand{\arraystretch}{1.3} 
\begin{tabular}{lllllr}
\toprule
 &  &  & \textbf{Release Type 1 Mean} & \textbf{Release Type 2 Mean} & \textbf{pvalue} \\
\textbf{Prog. Lang.} & \textbf{Release Type 1} & \textbf{Release Type 2} &  &  &  \\
\midrule
JavaScript & major & minor & 4.71 ($\pm$1.10) & 4.14 ($\pm$1.00) & $<$0.0001 \\
 &  & patch & 4.71 ($\pm$1.10) & 3.39 ($\pm$1.17) & $<$0.0001 \\
 & minor & patch & 4.14 ($\pm$1.00) & 3.39 ($\pm$1.17) & $<$0.0001 \\
\midrule
Python & major & minor & 5.34 ($\pm$1.06) & 4.76 ($\pm$1.02) & $<$0.0001 \\
 &  & patch & 5.34 ($\pm$1.06) & 4.17 ($\pm$1.15) & $<$0.0001 \\
 & minor & patch & 4.76 ($\pm$1.02) & 4.17 ($\pm$1.15) & $<$0.0001 \\
\midrule
Ruby & major & minor & 5.09 ($\pm$1.05) & 4.55 ($\pm$0.87) & $<$0.0001 \\
 &  & patch & 5.09 ($\pm$1.05) & 3.96 ($\pm$1.03) & $<$0.0001 \\
 & minor & patch & 4.55 ($\pm$0.87) & 3.96 ($\pm$1.03) & $<$0.0001\\
\bottomrule
\end{tabular}
\end{table}

However, a different trend is observed when examining the relationship between this assessed technical complexity and downstream adoption, measured by the log difference in dependents one year post-release. As shown in Figure \ref{fig:log-diff-dependents-release-complexities}, each programming language's distributions of log differences across complexity ratings (1 to 7) indicate limited correlation. Spearman's rank correlation coefficients for log-difference in dependents versus complexity rating are low for JavaScript (statistic = 0.015, p = 0.220) and Python (statistic = 0.024, p = 0.118), indicating no significant relationship in these ecosystems. For Ruby, a weak negative correlation is present (statistic = -0.057, p = 0.024), suggesting that higher complexity ratings are associated with slightly lower dependent growth. This general absence of a strong or consistent positive correlation indicates that the internal technical challenges or scale of changes involved in a release, from a maintainer's perspective, do not appear to be a key factor influencing whether end-users upgrade or integrate a new package version.

The combined insights from these analyses offer a dual perspective. Maintainers appear to align their internal development efforts with semantic versioning conventions, producing releases whose technical complexity generally corresponds to their declared version type (Major, Minor, or Patch). However, the limited correlation between release complexity and dependent growth indicates that downstream users' decisions to adopt new package versions are likely driven by factors other than the maintainers' development effort. This implies that end-users prioritize external factors such as new functionality, bug fixes, or perceived value rather than the intrinsic difficulty of the release when considering an upgrade.

\section{Discussion} \label{sec:discussion}

\subsection{Theoretical Implications}

\subsubsection{Making Visible OSS Innovation and Its Nuances}
Our proposed measure of software innovation helps closing a critical gap in the innovation literature, which has largely overlooked open-source software (OSS) outputs. The distinct downstream adoption patterns tied to semantic versioning—major, minor, and patch releases—offer a compelling illustration of \textit{incremental versus radical innovation} within a software context. This distinction is especially critical for OSS, whose contributions span small, iterative improvements to significant, transformative updates.

Crucially, our findings demonstrate that the impact of a release type on dependency growth is not uniform. Instead, it is \textit{highly contingent on the specific ecosystem (e.g., JavaScript, Python, Ruby), the package's pre-existing popularity, and its version series maturity (e.g., zero-version, one-version, two-plus-version)}. For instance, while minor releases may show consistent benefits across some ecosystems and package sizes, major or patch releases might exhibit varying impacts depending on ecosystem culture or baseline popularity. This highlights that "major" does not universally equate to "most impactful" in terms of rapid adoption.

Furthermore, in many emerging countries, measuring innovation is challenging due to limited data \cite{marins_2008} or a lack of incentives to patent innovations \cite{böhmeckedörries2023}. OSS innovation, with its cost-effectiveness and accessibility, offers substantial potential in these regions to drive technological advancement without relying on expensive proprietary solutions. Since traditional innovation metrics often fail to capture OSS's impact in these areas, our work provides an initial step toward developing new measurement methods. This paves the way for future research that could enable country-level analyses and more accurately assess OSS innovation, particularly where conventional metrics are inadequate.

\subsubsection{Complementing Innovation Indicators with Granular, Cost-Effective, and Context-Aware Measurement}
Established innovation indicators (discussed in \S~\ref{sec:relatedwork}) face significant limitations, particularly regarding timeliness and granularity. The Oslo Manual \cite{oecd_oslo_2018} highlights the challenge of measuring radical innovations due to delays in their effects, which can obscure the real-time dynamics of innovation. In contrast, our proposed measure enables more rapid evaluations of innovation—radical or incremental—even with a one-year window to validate trends in dependents. Unlike patent applications, which are published only after an 18-month period from the earliest filing date, our methodology facilitates analyzing how innovations are utilized, providing \textit{citation visibility} akin to scientific publications and practical implications similar to patents. By focusing on software innovation, our approach can also address the insufficient sectoral and technological granularity of traditional metrics (e.g., with a focus on the ICT sector). Moreover, conventional indicators often incur high data collection costs, especially at scale. By leveraging accessible data from developer platforms such as GitHub, our methodology offers scalable access to data without substantial financial investment. Additionally, our approach employs established criteria for assessing software innovation, with few qualitative aspects compared to, for example, survey-based innovation indicators. These qualitative aspects primarily relate to the documentation of release notes and the rigor of semantic versioning applied by contributors.

\subsubsection{Transparency and Visibility of Individual Contributions in the Innovation Ecosystem}
Our measure of software innovation enhances transparency in contributions: specific lines of code within an innovative package can be traced back to their creators. While not extensively explored in this paper, this visibility could support further work to evaluate individual contributions to innovation over time and their relation to trends in academic publishing and other innovative fields. Building on this, \textit{social network analysis (SNA)} presents an important avenue for future research in understanding the dynamics of OSS innovation ecosystems. Widely applied in innovation economics, SNA can offer a more in-depth exploration of how knowledge and innovations circulate within networks \cite{CantnerandGraf2006}. As noted in the literature review, approaches to network analysis of OSS networks have already been undertaken (see, e.g., \cite{banks2022measuring} or \cite{wiggins_heartbeat_2009}). However, by incorporating dependency data and differentiating inventive steps through semantic versioning, SNA could provide a more granular map of interactions between contributors in OSS ecosystems and contribute to our understanding of how knowledge and innovation circulate within OSS projects. For example, this approach could reveal how collaborative efforts shape innovation, identifying key nodes or bottlenecks and informing policies aimed at fostering more effective knowledge exchange and innovation flow.

\subsubsection{Potential Self-Fulfilling Prophecy in Release Labeling and Innovation Adoption}
Our measurements may demonstrate a \textit{self-fulfilling prophecy} that reflects the marketing of innovation. We find that the type of release is a strong predictor of downstream adoption. While there is a weak positive correlation between semantic versioning and release complexity, as assessed by release notes, developers may primarily pay attention to major releases due to their label or public sharing of a release in channels not observed in the GitHub data. This creates a risk of \textit{endogeneity}: when a developer claims their new release is a major one, this assertion may spur downstream adoption based more on the label than the actual substance of the release.

This phenomenon parallels challenges in other innovation metrics, such as those for patents or R\&D expenditures. In the patent landscape, the sheer number of filings does not inherently correlate with the quality or impact of the innovations. A high volume of patents might mask a lack of substantive progress, while truly innovative ideas could be underrepresented if they are not effectively communicated or marketed \cite{oecd2013}. Similarly, with R\&D expenditures, what is labeled as R\&D can encompass a wide range of activities that do not necessarily lead to innovation; for instance, a significant portion of R\&D spending might be directed toward marketing or administrative tasks rather than genuine innovative processes. This variability complicates the standardization of R\&D as a measure of innovation \cite{oecd2013}.

In the context of our research, the label of a "major release" could similarly drive adoption regardless of the release's technical merit or innovative substance. This highlights the importance of considering how the communication surrounding releases can shape perceptions and behaviors in the OSS community. The influence of promotional language and activity may significantly affect how innovation is recognized and valued, potentially leading to a misrepresentation of the innovation landscape. Further research on this possible phenomenon with off-platform data would support related research tracking promotional language and activity associated with academic research \cite{promotionallanguage,branch2024controlled}.

\subsubsection{Supporting Comparisons of Software Language Innovation Ecosystems}
Finally, our measurements of software innovation can be used to create a baseline for expected adoption for major, minor, or patch versions within a particular ecosystem, package size, and version series set. This could support research on \textit{language evolution} and \textit{comparative analysis between programming language ecosystems}, building on our observed variation between JavaScript, Python, and Ruby in dependent growth one year after a release. Within a particular language, the expected baseline can also identify especially innovative or impactful releases early on based on their divergent growth trajectory. Future work could describe this innovative subset and compare such a measure of 'innovativeness' to existing approaches like package citation combination novelty \cite{fang2024novelty}.

\subsection{Practical Implications}

\subsubsection{Informing National Measurement and Policy}
Beyond the innovation literature, our measure of software innovation has practical implications for policymakers, practitioners, and researchers. Measurements may be aggregated by jurisdiction, as has been done to date with software development activities in the \href{https://innovationgraph.github.com/}{GitHub Innovation Graph}, and inform international comparisons of innovation \cite{gii}. Governments may look to the measure as one data point to inform the evaluation of national industrial strategies, technology-specific (e.g., AI) strategies, and other initiatives that aim to spur innovation.

For practitioners in the private sector, our approach can support the evaluation of investment decisions and outcomes. By analyzing software innovation metrics, organizations can gain insights into market trends, competitive positioning, and potential return on investment. This can enable more informed decision-making regarding research and development initiatives, partnerships, and resource allocation.

\subsubsection{Refining OSS Innovation Metrics with Contextual Normalization}
Researchers and practitioners focusing on OSS can leverage the proposed measure of software innovation to advance their understanding of how innovation unfolds within different OSS ecosystems. Previous literature \cite{blind_impact_2021} measuring the impact of OSS has primarily relied on commits as proxies for OSS activity, correlating these with broader economic and social outcomes. However, commits may not adequately capture innovation since they do not distinguish between incremental updates and substantial, innovative changes.

Our findings strongly suggest that future OSS innovation and adoption measures should incorporate \textit{contextual normalization schemes*}, akin to "Field Weighted Citation Impact" used in scientometrics \cite{purkayastha2019comparison}. As research impact is normalized by field and publication year, OSS innovation metrics should account for the specific ecosystem, package size, and version series. This ensures that the measured "impact" of a release (e.g., in terms of dependent growth) is evaluated relative to what is expected within its specific context, providing a more accurate and equitable assessment of its innovativeness.

For future research, we may consider the \textit{ratio of dependents to dependencies} as a measure of a repository's innovativeness, similar to the forward-to-backward citation ratio used in bibliometric and patent analysis \cite{dahlander2010open}. This approach aligns with the concept of revealed preference, suggesting that when developers choose to depend on a project rather than writing code from scratch, they signal that the dependency represents a non-trivial inventive step. Our preliminary exploration of the top 1,000 GitHub repositories, ranked by stars, forks, and issue authors, has shown a positive correlation between the dependent-dependency ratio and the number of dependents. This suggests that repositories with a high ratio of dependents to dependencies can be considered more innovative, as they exert a greater influence on the OSS ecosystem relative to their reliance on other projects.

Using semantic versioning as output units of innovation provides a clearer view of substantial changes in software development that could potentially have broader economic and technological impacts. While directly linking these releases to specific economic outcomes—such as increased demand for technical services, job creation, or shifts in industry productivity—remains subject to further research, this measure may serve as an early indicator of such impacts. For example, downstream adoption patterns following a major release could signal broader trends in the uptake of new technologies or the influence of OSS on industry innovation.

However, in this work, we chose not to rely on a project's ratio of dependents to dependencies as the primary unit for *individual release* innovation, as it appeared less helpful in comparing individual releases within a project due to infrequent significant dependency changes. Nonetheless, project-to-project comparisons using the dependent-dependency ratio remain promising for identifying innovative software. Future studies should explore these methods to connect the proposed OSS innovation metrics to broader economic or sectoral trends.

While our measure of OSS innovation is primarily designed to track innovation within the software development ecosystem itself, future research could investigate potential links to broader outcomes, such as industry adoption rates or collaborative dynamics across programming communities. In this sense, our log-difference measurements provide a foundation for more granular OSS innovation analysis without venturing into economic impact analysis.

\subsection{Limitations}
This work faces the following limitations. First, we analyze only the direct dependents of a given package. This undervalues packages that have become central to software ecosystems, where their direct dependencies are, in turn, relied upon widely. Future work could remedy this limitation by analyzing indirect dependencies, albeit at significantly higher computational costs.

Second, our analysis does not account for the role of software licenses in shaping innovation patterns. Different license types (e.g., GPL vs. Apache v2) may systematically influence developer choices and dependency relationships, as licenses can enable or constrain how packages build upon each other. Furthermore, we do not address the geographic dimension of software innovation diffusion, which could reveal important regional variations in OSS adoption and development patterns.

Third, mature software packages often have platforms for communicating and distributing releases that are not captured in GitHub data. For example, some version releases on GitHub may use release notes that link to further details off-platform. Thus, our method may not accurately account for the complexity of mature packages.

Fourth and relatedly, our analysis of release complexity examines only the notes describing the release. Further study could examine the code associated with each release directly as another, perhaps more direct, indicator of complexity. Fifth, we examine only three software languages. Future work could expand the language coverage of the measure. However, some languages, such as C and C++, do not have centralized package ecosystems or rely heavily on GitHub's Release feature, which undermines the universal application of our measure.

Finally, while using LLMs to assess semantic content offers novel analytical possibilities, this approach has inherent limitations. Recent research suggests that LLM responses can vary significantly even with identical prompts \cite{arlinghaus2024inductive, tai2024examination}. Future work could enhance reliability by implementing multiple prompt iterations and employing mean responses as final classifications alongside expanded human validation procedures.

\section{Conclusion}\label{sec:conclusion}

This paper introduced a method for measuring software innovation, complementing traditional indicators by analyzing GitHub data on package releases and dependency adoption. Our findings reveal that the impact of different release types (major, minor, patch) on dependency growth depends on factors like the programming language ecosystem, package popularity, and version series maturity. This means a "major" release isn't always the most impactful; its effect depends on its context. Interestingly, while maintainers generally align a release's technical complexity with its semantic versioning type, the complexity of a release shows limited direct correlation with its downstream dependency growth. This suggests that release type can act as a signaling mechanism, influencing adoption somewhat independently of the technical effort involved. For future research, this highlights the need for normalized approaches to measuring OSS innovation, similar to "Field Weighted Citation Impact," to account for these contextual nuances and provide a more accurate assessment of a release's impact.

\newpage
\bibliographystyle{ieeetr}  
\bibliography{refs}

\newpage
\appendix
\section{Complexity rating LLM prompt}\label{sec:appendix-prompt}

\texttt{
SYSTEM\_PROMPT = """
You are computer science grad student. As a computer science grad student, you are extremely knowledgeable in software development, development lifecycles, and release patterns. Further, you also know and use multiple programming languages and frameworks. Currently you are tasked with rating the complexity of a software release given the release notes. As this is a research project, you will be provided an annotation procedure and the example to annotate. Be sure to follow the procedure and always respond with the XML formatted rating information.
"""
}

\texttt{
USER\_PROMPT = """
\char35\char35 Task
}

\texttt{
Rate the complexity of a software release given the software release's release notes. Specifically, you should rate the "complexity to implement" (i.e. how difficult does the feature, or bug, or change seem to be to implement).
}

\texttt{
For your "complexity to implement" rating, take on the persona of a core developer for the library. That is, some things may be easier or harder in different languages and those differences should be taken into account.
}
\texttt{
I know you will do great! Just try your best!
}

\texttt{
\char35\char35\char35 Rating Scale for Complexity to Implement
}

\texttt{
Use the following scale and criteria to rate the complexity to implement:
}

\texttt{
- 1. Almost no changes.
}

\texttt{
If any, they may be purely for documentation, project administration, or very minor bugfixes such as a typo or a small formatting issue.
}

\texttt{
- 2. Very few changes.
}

\texttt{
They may involve a new feature or a minor bugfix. But the changes are entirely minor. The changes are so small that they may not require any new documentation outside of a very brief mention in the release notes.
}

\texttt{
- 3. A few changes.
}

\texttt{
Changes involve some basic understanding of the library. They may involve a new feature or a minor bugfix. But the feature itself shouldn't be major. For example, it may be the addition of a new parameter to a function, or a new method to a class. It may require some new documentation but not a lot.
}

\texttt{
- 4. A small number of changes.
}

\texttt{
Changes involve some moderate level of understanding the library. They may involve a new major feature or a major bugfix. They may require some refactoring or changes to existing code but that isn't the main focus of the release. They likely require some additional documentation to announce the new feature, bugfix, or new behavior.
}

\texttt{
- 5. A moderate number of changes.
}

\texttt{
Changes involve a decently-high level of understanding the library. This may include multiple new features, major bugfixes or changes, or a moderate amount of refactoring. They should require extensive documentation to announce the new features, bugfixes, or new behavior.
}

\texttt{
- 6. A large number of changes.
}

\texttt{
Changes involve a high level of understanding the library. There are multiple new features, major bugfixes or changes, and/or a large amount of refactoring. Each change may be interacting with multiple systems or modules of the library or tool.
}

\texttt{
- 7. Extensive changes.
}

\texttt{
The release includes new features, refactoring. Complex interactions between multiple systems. To implement these changes would require extensive knowledge of the whole library to fully understand the effects of each change. Further, it would require extensive testing to ensure that the changes are correct and do not break existing functionality. This also requires extensive documentation to explain the changes to users.
}

\texttt{
\char35\char35 Input Structure}
\begin{verbatim}
You will be provided with an XML object with the following structure:
<release-notes-information>
    <repository-name>...</repository-name>
    <repository-description>...</repository-description>
    <repository-topics>...</repository-topics>
    <repository-language>...</repository-language>
    <release-notes>...</release-notes>
</release-notes-information>
\end{verbatim}

\texttt{\char35\char35 Response Structure
}

\texttt{
- `required-skills`: A list of short (less than a sentence) semi-colon separated notes, of the skills required to implement the changes in the release notes. I.e. what knowledge (e.g. asynchrony, data structures, etc.) would be required to implement the changes.
}

\texttt{
- `reasoning`: A list of short (less than a sentence) semi-colon separated notes, that justify the rating you are providing.
}

\texttt{
- `complexity-rating`: The complexity rating you are providing. This should be a number between 1 and 7, inclusive, where 1 is "very low complexity" and 7 is "very high complexity". Always try to rate the release on the scale from 1 to 7, however, if there is almost no information in the release notes, rate the complexity as "null".
}

\texttt{Your response should have the following structure:}
\begin{verbatim}
<rating-response>
    <required-skills>...</required-skills>
    <reasoning>...</reasoning>
    <complexity-rating>...</complexity-rating>
</classification-response>
\end{verbatim}

\texttt{Provide only the XML response, without any additional text or formatting.}

\texttt{
\char35\char35 Release Information
}
\begin{verbatim}
<release-notes-information>
    <repository-name>{repo_name}</repository-name>
    <repository-description>{repo_description}</repository-description>
    <repository-topics>{repo_topics}</repository-topics>
    <repository-language>{repo_language}</repository-language>
    <release-notes>{release_notes}</release-notes>
</release-notes-information>
\end{verbatim}

\texttt{
\char35\char35 Rating Response
"""
}

\newpage
\section{One-Year 90 Day Timepoint Distributions of Log-Differences of Dependents for Pre-Existing Package Utilization Bins}\label{sec:appendix-one-year-timepoints-boxplots}

\begin{figure}[htbp]
    \centering
    \includegraphics[width=\textwidth]{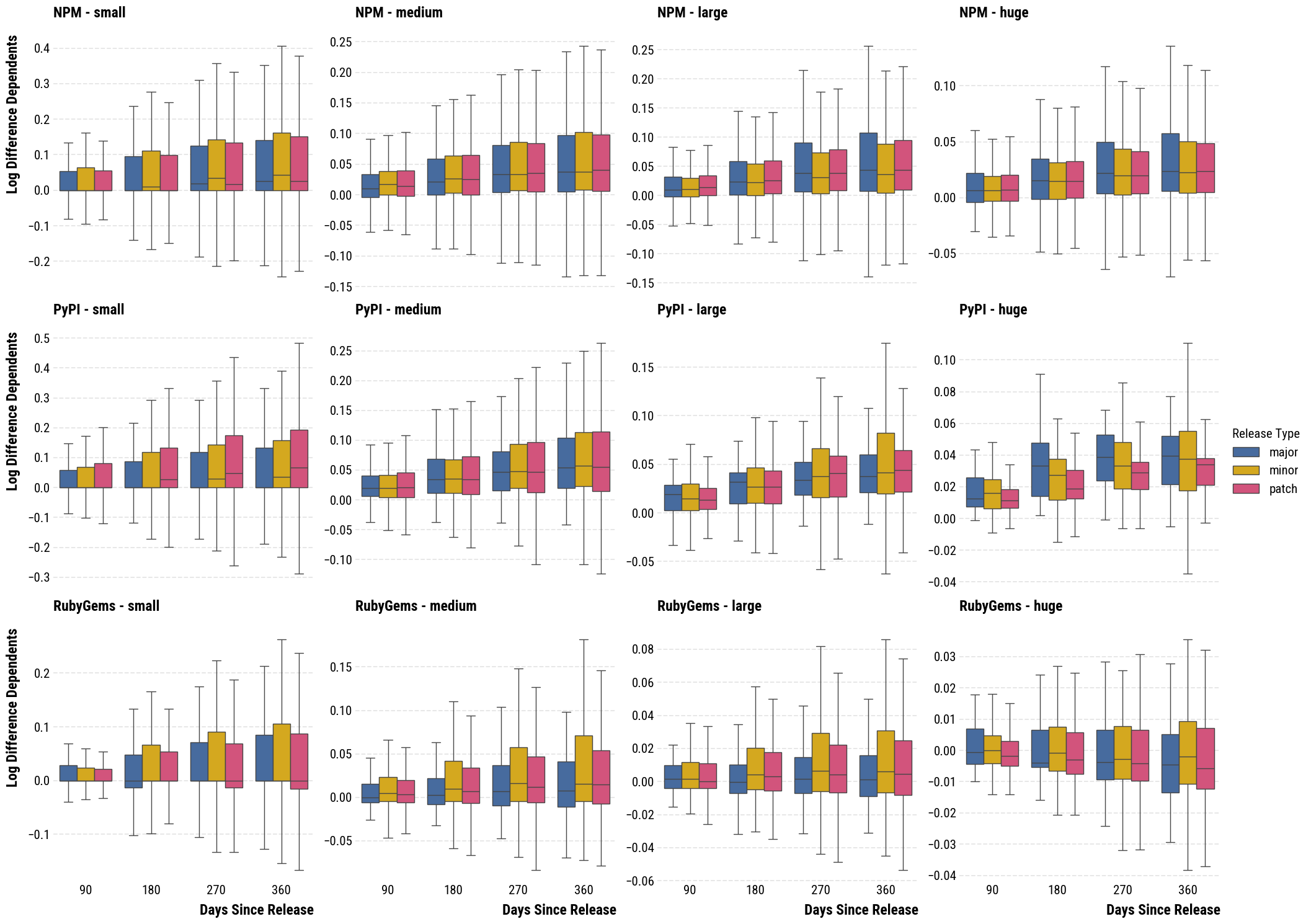}
    \caption{Distributions of log-difference dependents split by Ecosystem (NPM, PyPI, and RubyGems), pre-release package size (small, medium, large, and huge), release type (major, minor, or patch), and over multiple 90-day timepoints for one year after release.}
\end{figure}

\newpage
\section{Two-Year 180 Day Timepoint Distributions of Log-Differences of Dependents for Pre-Existing Package Utilization Bins}\label{sec:appendix-two-year-timepoints-boxplots}

\begin{figure}[htbp]
    \centering
    \includegraphics[width=\textwidth]{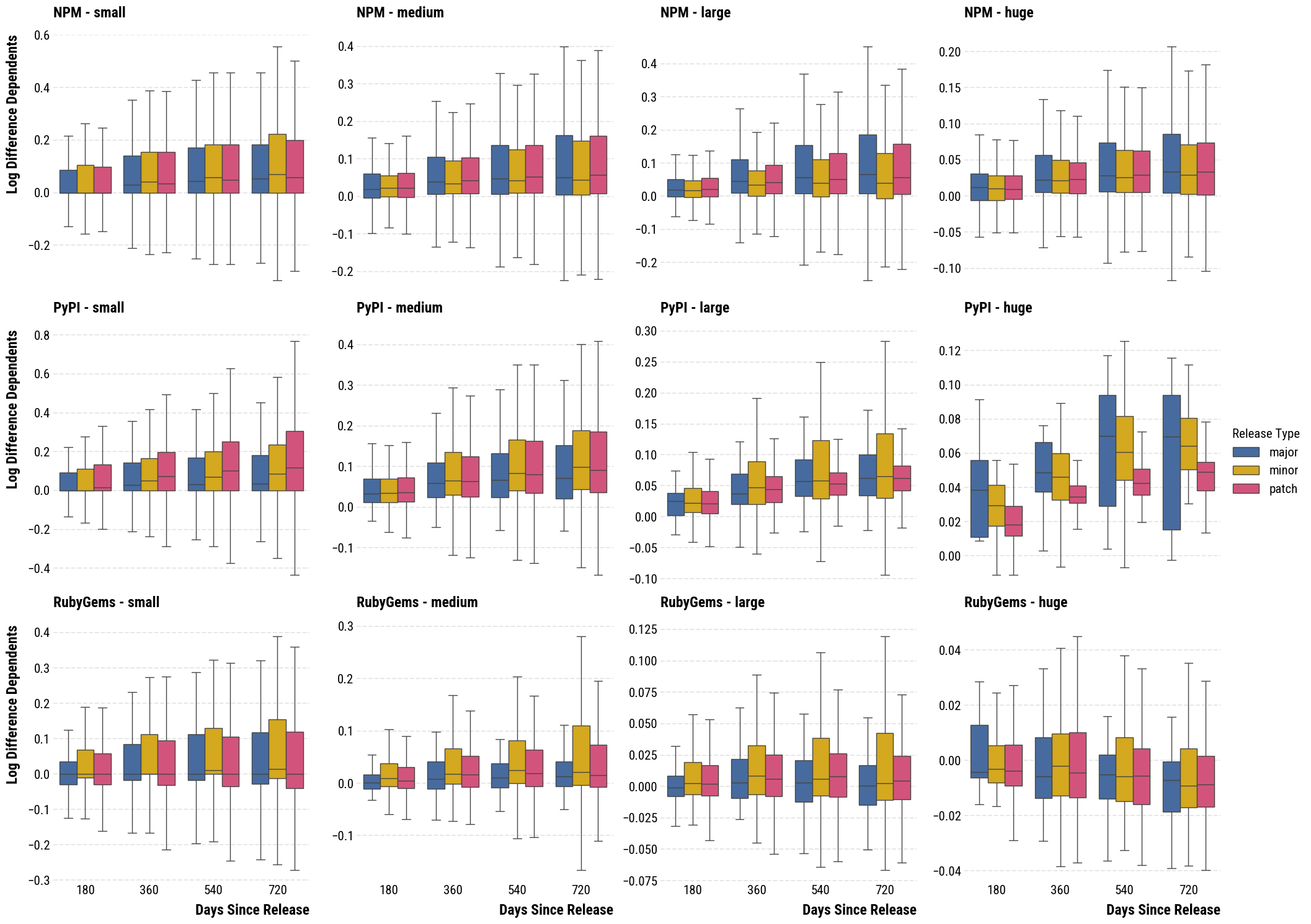}
    \caption{Distributions of log-difference dependents split by Ecosystem (NPM, PyPI, and RubyGems), pre-release package size (small, medium, large, and huge), release type (major, minor, or patch), and over multiple 180-day timepoints for two years after release.}
\end{figure}

\newpage
\section{One-Year 90 Day Timepoint Distributions of Log-Differences of Dependents for Version Series Bins}\label{sec:appendix-one-year-timepoints-boxplots-ver-series}

\begin{figure}[htbp]
    \centering
    \includegraphics[width=\textwidth]{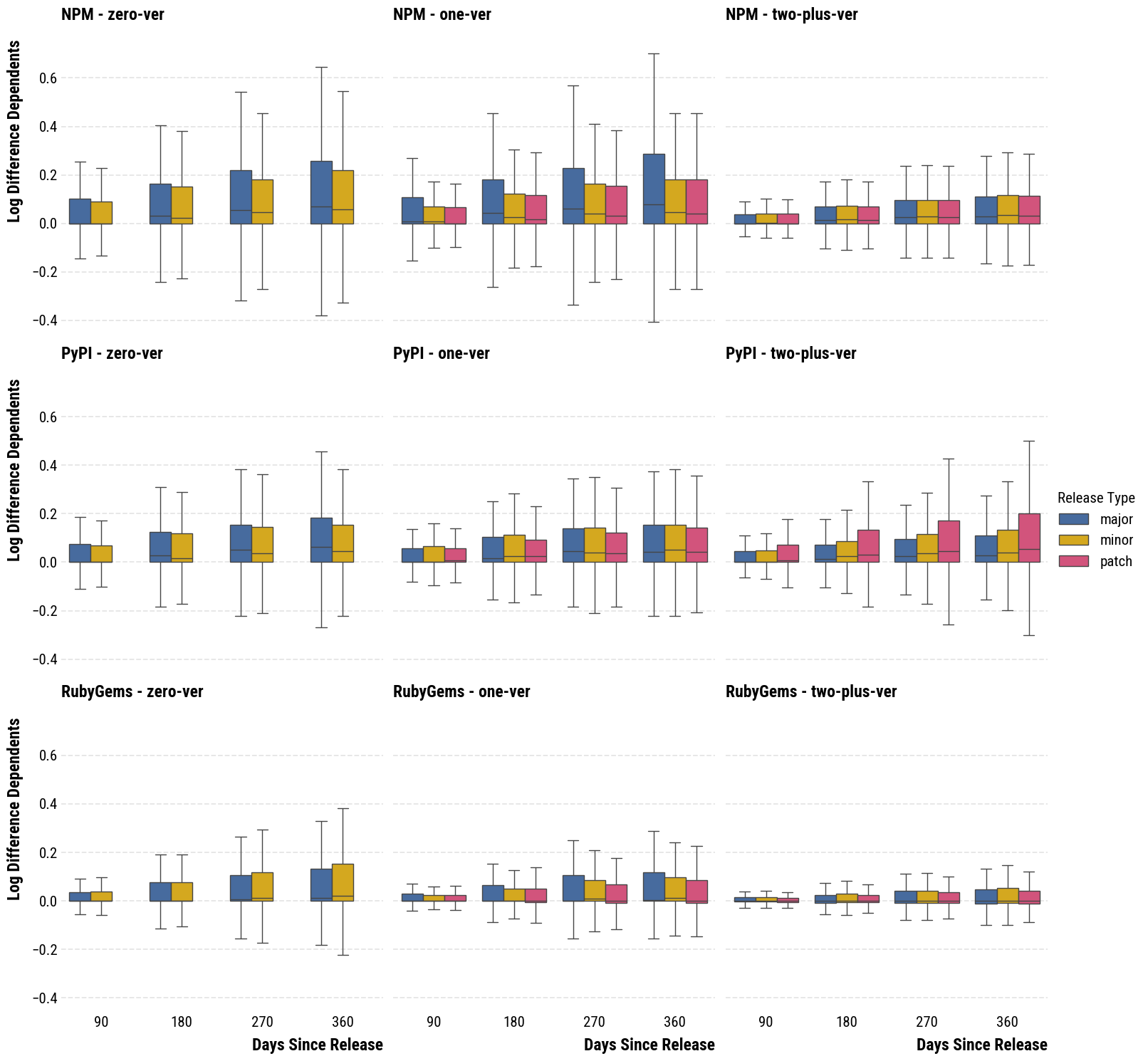}
    \caption{Distributions of log-difference dependents split by Ecosystem (NPM, PyPI, and RubyGems), version series (zero-version, one-version, and two-plus-version), release type (major, minor, or patch), and over multiple 90-day timepoints for one year after release.}
\end{figure}

\newpage
\section{Two-Year 180 Day Timepoint Distributions of Log-Differences of Dependents for Version Series Bins}\label{sec:appendix-two-year-timepoints-boxplots-ver-series}

\begin{figure}[htbp]
    \centering
    \includegraphics[width=\textwidth]{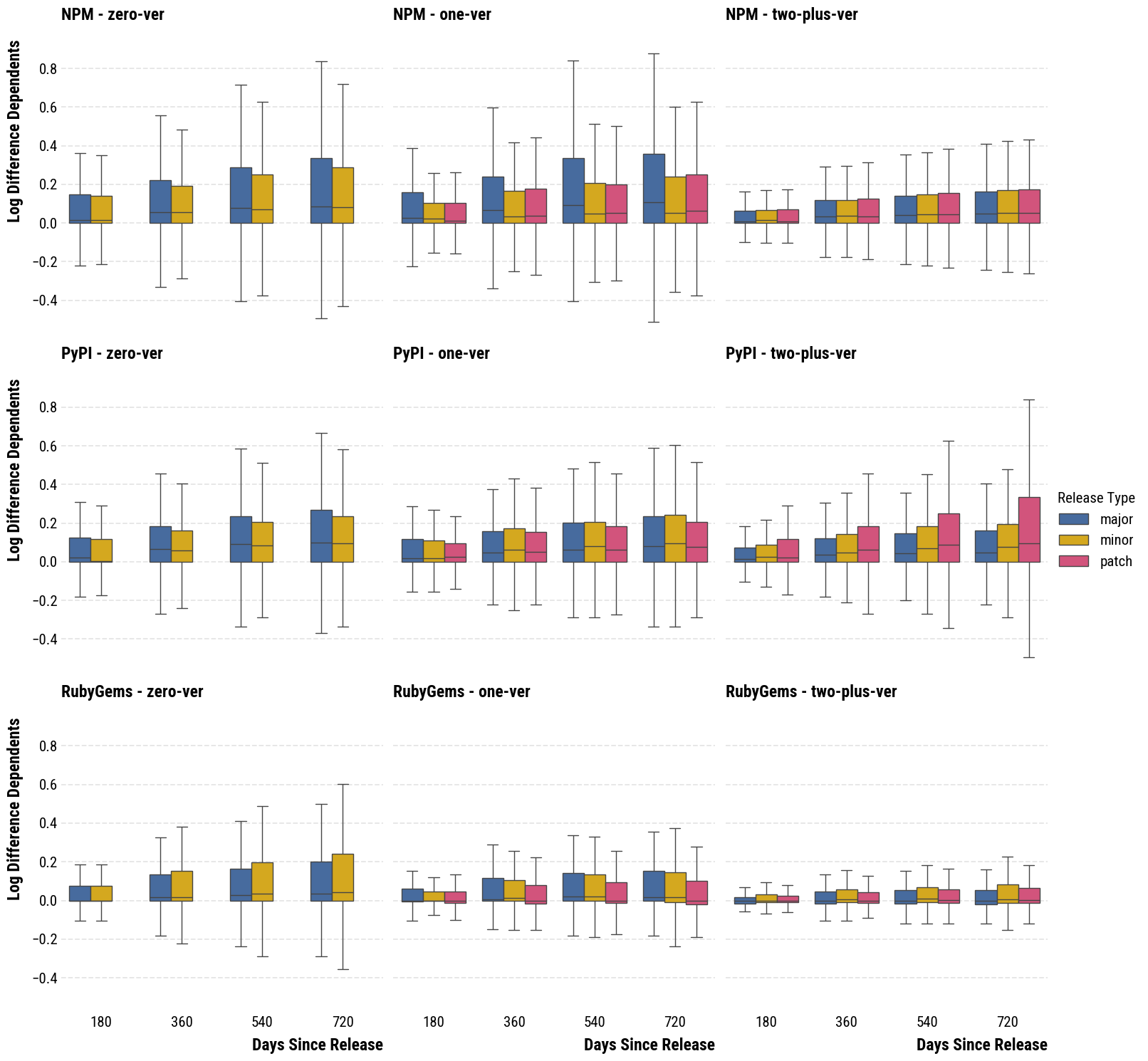}
    \caption{Distributions of log-difference dependents split by Ecosystem (NPM, PyPI, and RubyGems), version series (zero-version, one-version, and two-plus-version), release type (major, minor, or patch), and over multiple 180-day timepoints for two years after release.}
\end{figure}



\end{document}